\documentclass[aps,reprint,prb,amsmath,amssymb]{revtex4-1}
\usepackage{graphics}
\usepackage{graphicx}
\usepackage{dcolumn} 
\usepackage{bm}
\usepackage{epsfig}

\begin{document}
\title{Uniaxial contribution to the magnetic anisotropy of La$_{0.67}$Sr$_{0.33}$MnO$_3$ thin films induced by orthorhombic crystal structure}
\author{Hans Boschker}
\author{Mercy Mathews}
\author{Peter Brinks}
\author{Evert Houwman}
 \email{e.p.houwman@utwente.nl}
\author{Gertjan Koster}
\author{Dave H. A. Blank}
\author{Guus Rijnders}
\affiliation{Faculty of Science and Technology and MESA$^+$ Institute for Nanotechnology, University of Twente, 7500 AE, Enschede, The Netherlands}

\author{Arturas Vailionis}
\affiliation{Geballe Laboratory for Advanced Materials, Stanford University, Stanford, California 94305, USA }

\date{\today}

\begin{abstract}
La$_{0.67}$Sr$_{33}$MnO$_3$ (LSMO) thin films under compressive strain have an orthorhombic symmetry with (1$\overline{1}$0)$_\textrm{o}$ and (001)$_\textrm{o}$ in-plane orientations. (The subscript o denotes the orthorhombic symmetry.) Here, we grew LSMO on cubic (LaAlO$_3$)$_{0.3}$-(Sr$_2$AlTaO$_6$)$_{0.7}$ (LSAT) substrates and observed a uniaxial contribution to the magnetic anisotropy which is related to the orthorhombic crystal structure. Since the lattice mismatch is equal in the two directions, the general understanding of anisotropy in LSMO, which relates the uniaxial anisotropy to differences in strain, cannot explain the results. These findings suggest that the oxygen octahedra rotations associated with the orthorhombic structure, possibly resulting in different Mn-O-Mn bond angles and therefore a change in magnetic coupling between the [1$\overline{1}$0]$_\textrm{o}$ and [001]$_\textrm{o}$ directions, determine the anisotropy. We expect these findings to lead to a better understanding of the microscopic origin of the magnetocrystalline anisotropy in LSMO.

\end{abstract}

\maketitle

\section{Introduction}

The perovskite oxide La$_{1-x}$A$_{x}$MnO$_3$ (A=Ca, Ba, Sr) has initiated a substantial body of research due to its colossal magnetoresistance \cite{Helmolt1993, Jin1994}. Extensive theoretical studies and experimental investigations utilizing La$_{1-x}$A$_{x}$MnO$_3$ perovskites in bulk form revealed a strong coupling between lattice distortions and magnetism, which substantially modify magnetic properties such as magnetoresistance and Curie temperature \cite{Moritomo1995, Hwang1995prb}. La$_{0.67}$Sr$_{33}$MnO$_3$ (LSMO) has the highest Curie temperature (370K) and a 100\% spin polarization \cite{Park1998, Bowen2003}. LSMO can be coherently grown on a range of commercially available perovskite substrates, such as e.g. NdGaO$_3$ (NGO) and SrTiO$_3$ (STO). The epitaxy stabilizes a different crystal structure which modifies the magnetic properties. Especially magnetic anisotropy is shown to be very sensitive to the LSMO crystal structure \cite{Kwon1997, Suzuki1998, Steenbeck1999, Tsui2000, Desfeux2001, Dho2003, Mathews2005}. When anisotropic strain is applied to the LSMO the magnetocrystalline anisotropy becomes strongly uniaxial \cite{Boschker2009, Mathews2010}, which is a useful tool to tailor the magnetic properties for device applications.

In the case of isotropic tensile strain, e.g. tetragonal LSMO thin films on cubic STO (001)$_\textrm{c}$ substrates, the magnetocrystalline anisotropy is biaxial with easy axes aligned with the $<$110$>$$_\textrm{pc}$ lattice directions \cite{Steenbeck1999, Tsui2000}. (We use subscript c, pc, o and t for cubic, pseudocubic, orthorhombic and tetragonal crystal structures, respectively.) Next to the magnetocrystalline anisotropy a uniaxial anisotropy is present as well, which is stepedge induced \cite{Wang2003, Mathews2005}. Here we investigate the case of isotropic compressive strain, which can be realized with LSMO thin films on the cubic (LaAlO$_3$)$_{0.3}$-(Sr$_2$AlTaO$_6$)$_{0.7}$ (LSAT) (001)$_\textrm{c}$ substrate. LSMO thin films under compressive strain adopt an orthorhombic crystal structure \cite{Vailionis2008, Vailionis2009}, which is characterized by the presence of oxygen octahedra rotations around all three pseudocubic crystal axes. As the magnetic coupling depends on the Mn-O-Mn bond angle \cite{Hwang1995, Radaelli1997}, it is an interesting question whether the magnetic properties are anisotropic in the different orthorhombic directions. Note that for another case, orthorhombic LSMO grown on NGO (110)$_\textrm{o}$ the difference in lattice mismatch between the two in-plane directions determines the anisotropy \cite{Boschker2009}, so this system is not suitable to study the effect of the orthorhombicity on the magnetic properties. For LSMO films grown on NGO (110)$_\textrm{o}$ the [1$\overline{1}$0]$_\textrm{o}$ lattice direction is subjected to less compressive strain than the [001]$_\textrm{o}$ lattice direction and is therefore the easy axis due to the strain anisotropy. For LSMO films grown on LSAT the lattice mismatch is equal and the anisotropy is due to the intrinsic anisotropy of the orthorhombic crystal structure between the [1$\overline{1}$0]$_\textrm{o}$ and [001]$_\textrm{o}$ lattice directions.

Here, we show that LSMO thin films can be grown coherently and untwinned on LSAT substrates and that the orthorhombicity induces anisotropic magnetic properties. Next to a biaxial component of the magnetic anisotropy, we observed a uniaxial component to the anisotropy which is aligned with the principal crystal directions and became more pronounced for increasing film thickness. We found no correlation between the uniaxial anisotropy and the stepedge direction. We obtained twinned samples, by growth on surfaces with reduced crystalline quality, for which the uniaxial anisotropy was reduced. Therefore we conclude that the uniaxial anisotropy is caused by the orthorhombic crystal structure.

\section{Samples and substrate preparation}
\begin{figure}
\centering
\includegraphics*[width=8cm]{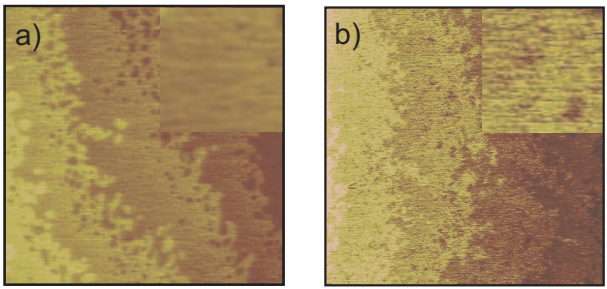}
\caption{(Color online) Surface analysis of the LSAT substrate by atomic force microscopy. a) after annealing at 1050$^\circ$C for 12 hours. b) after annealing at 950$^\circ$C for 1 hour,  The images are 5 by 5 $\mu$m and the color scale is 2 nm. The insets show a close-up of the roughness of the terraces.}
\label{substrates}
\end{figure}

The as-received LSAT substrates were cleaned with acetone and ethanol before they were subjected to an anneal treatment. Two anneal treatments were used to obtain respectively surfaces with smooth terraces and surfaces with sub unit cell roughness on the terraces. The first treatment consisted of an annealing step at 1050$^\circ$C for 12 hour in 1 bar of O$_2$ gas pressure. For the second treatment both the anneal time and temperature were decreased to 1 hours and 950$^\circ$C respectively. The surfaces were characterized with atomic force microscopy (AFM). Typical results are shown in figure~\ref{substrates}. For the substrates subjected to the first anneal treatment a step and terrace structure with~4\AA~(a single unit cell) step height was observed. The stepedges were not straight but meandering and~4\AA~deep holes are observed near the stepedges. Note that the miscut of these substrates is very small, approximately 0.02$^\circ$, giving a terrace width of more than 1 $\mu$m. Between the stepedges areas with atomically smooth morphology were observed. The substrates subjected to the second treatment show terraces with reduced crystalline quality, but still single unit cell step heights.

LSMO thin films were grown on the LSAT (001) substrates by pulsed laser deposition (PLD) from a stoichiometric target in an oxygen background pressure of 0.35 mbar with a laser fluence of 3 J/cm$^2$ and at a substrate temperature of 750$^\circ$C. After LSMO deposition, the films were cooled to room temperature at a rate of 10$^\circ$C/min in a 1 bar pure oxygen atmosphere. The growth settings were previously optimized and were identical to the ones used for LSMO films on other substrates \cite{MathewsPHD, Boschker2009}.

In this paper four samples are described, see table~\ref{table}. Sample U12 and U40 were grown on substrates with a smooth surface and have a thickness of 12 and 40 nm respectively. Samples T29 and T50 were grown on substrates with terraces with reduced crystalline quality and are respectively 29 and 50 nm thick. (The sample labels consist of either the letter T or U for twinned/untwinned and a number which indicates the sample thickness.) The sample thicknesses were measured with x-ray reflectivity measurements. AFM measurements (not shown) revealed surfaces of the thin film where the morphology of the substrate was still visible. The Curie temperature of the films was larger than 350 K (350 K was the measurement limit of the vibrating sample magnetometer) and did not depend on film thickness and the twinning, as discussed in the next section, of the films.

\begin{table*}
\begin{center}
\begin{tabular}{|c|c|c|c|c|c|c|c|}
\hline
&&&&&&&\\[-1.5ex]
Sample&  Thickness &Substrate & Crystal &Satellites &$\phi_\textrm{easy}$($^\circ$) & $k_\textrm{u}/k_1$ & $k_\textrm{u}$ \\
&(nm)&surface&&&&& (J/m$^3$)\\
&&&&&&&\\[-1.5ex]
\hline
&&&&&&&\\[-1.5ex]
U12 &12&smooth&untwinned&not observed	&	40$\pm$1  &	0.18 $\pm$0.05 &  110$\pm$30\\
U40	&40&smooth&untwinned&along [001]$_\textrm{o}$&	12$\pm$1  &	0.91 $\pm$0.03 &	540$\pm$5\\
T29	&29&rough&twinned&both directions&	42$\pm$1  &	0.1 $\pm$0.05    &  60$\pm$30\\
T50	&50&rough&twinned&both directions&	31$\pm$1  &	0.48 $\pm$0.03   &  290$\pm$25\\
\hline
\end{tabular}
\caption{The ratio between the anisotropy constants for the various samples. The angle of the easy axis is obtained from the fitting procedure and the ratio between the anisotropy constants is calculated with equation \ref{phieasy}.}
\label{table}
\end{center}
\end{table*}

\section{Structural characterization}
\begin{figure}
\centering
\includegraphics*[width=8cm]{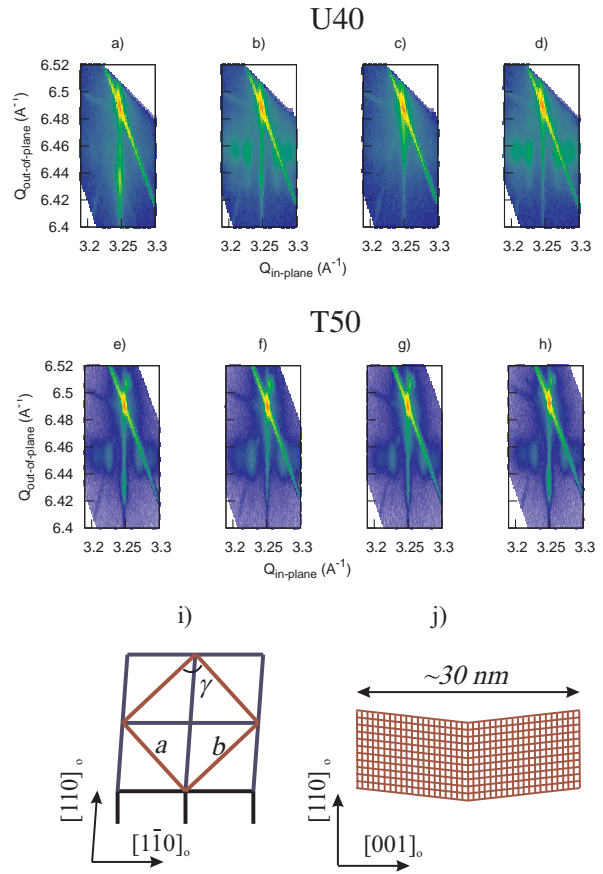}
\caption{[Color online] Top panel: Reciprocal space maps around the a) (204)$_\textrm{c}$, b) (024)$_\textrm{c}$, c) ($\overline{2}$04)$_\textrm{c}$ and d) (0$\overline{2}$4)$_\textrm{c}$ LSAT reflections of the 40 nm thick sample grown on a smooth substrate (sample U40). In a) and c) the dissimilar spacing of the (260)$_\textrm{o}$ and (620)$_\textrm{o}$ LSMO reflections is clearly observed while in b) and d) satellites are present next to the (444)$_\textrm{o}$ and (44$\overline{4}$)$_\textrm{o}$ LSMO reflections. Middle panel: Reciprocal space maps around the e) (204)$_\textrm{c}$, f) (024)$_\textrm{c}$, g) ($\overline{2}$04)$_\textrm{c}$ and h) (0$\overline{2}$4)$_\textrm{c}$ LSAT reflections of the 50 nm thick sample grown on a rough substrate (sample T50). The satellites are present in all maps and both e) and h) show intensity at the position for the (260)$_\textrm{o}$ LSMO reflection. Therefore this sample is twinned. Bottom panel: Schematic of the real space crystal structure of an untwinned sample viewed along i) [001]$_\textrm{o}$ and j) [1$\overline{1}$0]$_\textrm{o}$.}
\label{xrd}
\end{figure}

The top panel of figure \ref{xrd} shows reciprocal space maps of LSMO and LSAT around the (204)$_\textrm{c}$, (024)$_\textrm{c}$, ($\overline{2}$04)$_\textrm{c}$ and (0$\overline{2}$4)$_\textrm{c}$ LSAT reflections. These results were obtained from sample U40. The LSMO has a slightly distorted orthorhombic (monoclinic) unit cell with (110)$_\textrm{o}$ out-of-plane orientation and (1$\overline{1}$0)$_\textrm{o}$ and (001)$_\textrm{o}$ in-plane orientations. The orthorhombicity can be deduced from figures \ref{xrd}a and \ref{xrd}c which show a difference in lattice spacing for the (260)$_\textrm{o}$ and (620)$_\textrm{o}$ LSMO reflections, which represent a dissimilarity between the LSMO a and b lattice parameters. The lattice parameters are as follows: a=5.47$\pm$0.01\AA, b=5.51$\pm$0.01\AA,
c=7.74$\pm$0.01\AA, $\alpha$=90$\pm$0.1$^\circ$, $\beta$=90$\pm$0.1$^\circ$ and $\gamma$=89.6$\pm$0.1$^\circ$. 

Next to the LSMO (444)$_\textrm{o}$ and (44$\overline{4}$)$_\textrm{o}$ reflections satellites are observed. In a previous paper \cite{Vailionis2009} we have discussed the orthorhombic crystal structure and the presence of satellite peaks in the [001]$_\textrm{o}$ direction for LSMO grown on NGO (110)$_\textrm{o}$ substrates. The satellites result from periodic lattice modulations which partially relieve the applied strain \cite{Vigliante2001, Farag2005, Gebhardt2007}. As the LSMO crystal structure can easily relieve strain in the [1$\overline{1}$0]$_\textrm{o}$ lattice direction with a change in the $\gamma$ angle, the lattice modulations are only present in the [001]$_\textrm{o}$ direction. We conclude that LSMO films on LSAT behave similarly as LSMO on NGO.

The panels e-h) of figure \ref{xrd} show the same reciprocal space maps, but obtained for sample T50. Dissimilar lattice spacings are observed in both in-plane directions and the satellite peaks are visible in all reciprocal space maps. Zhou et al. observed similar satellite peaks around reflections of LSMO on LSAT and attributed the satellites to an in-plane superlattice with alternating [1$\overline{1}$0]$_\textrm{o}$ and [$\overline{1}1$0]$_\textrm{o}$ orientations \cite{Zhou2007}. This cannot be the case for our samples as the [$\overline{1}1$0]$_\textrm{o}$ in-plane orientation would result in reciprocal space maps with both the (620)$_\textrm{o}$ and the (260)$_\textrm{o}$ peak visible in the same plot. In e.g. figure \ref{xrd}f this is clearly not the case. Together with the presence of the satellites in figure \ref{xrd}b and \ref{xrd}d only in the [001]$_\textrm{o}$ direction, we conclude that sample D is twinned in domains with different [1$\overline{1}$0]$_\textrm{o}$ and [001]$_\textrm{o}$ orientations.

For the two thinner samples we observed the same behavior as the thicker samples with equal substrate treatment (not shown). Sample U12 was untwinned as concluded from the positions of the Bragg reflections, but here no satellites could be observed. Sample T29 was twinned with satellites in both directions in reciprocal space. 

To explain the absence of twinning in the samples with smooth surfaces we compare our results to the more widely studied SrRuO$_3$ (SRO) thin films on STO substrates. SRO is orthorhombic and single domain films can be realized by growth on smooth vicinal substrates with stepedges approximately aligned with the main crystal axis. In that case the [001]$_\textrm{o}$ lattice directions aligns with the stepedge direction \cite{Gan1997}. This has been explained in three different ways. Gan et al. suggested that single domain growth is related to the observed stepflow growth mode \cite{Gan1997}. Maria et al. suggested that stepedge strain is the dominant mechanism as the films are not orthorhombic at deposition temperature \cite{Maria2000}. Finally Vailionis et al. suggested that the films are tetragonal at deposition temperature and that the [001]$_\textrm{t}$ lattice direction aligns preferentially with the stepedges \cite{Vailionis2007}. In contrast to SRO the LSMO films have a preferred orientation with the [001]$_\textrm{o}$ lattice direction aligned perpendicular to the stepedges and does not grow in stepflow mode. Therefore the explanation by Maria et al. is the most likely candidate to explain the single domain growth. During cooldown orthorhombic domains nucleate at the stepedges and the stepedge strain favors octahedra buckling in one direction and the domains continue to grow across the terraces resulting in a single domain film. For the samples with a rough surface the orthorhombic domains can nucleate at defects at the substrate surface and no preferential orientation exists, resulting in twinned films. 

In summary, growth on a relatively smooth surface results in untwinned LSMO films, while growth on terraces with reduced crystalline quality results in twinned LSMO films. We used the difference in magnetic properties between twinned and untwinned films to identify the contribution from the orthorhombicity as it should be reduced for the twinned samples.

\section{Magnetic characterization}
\begin{figure}
\centering
\includegraphics*[width=9cm]{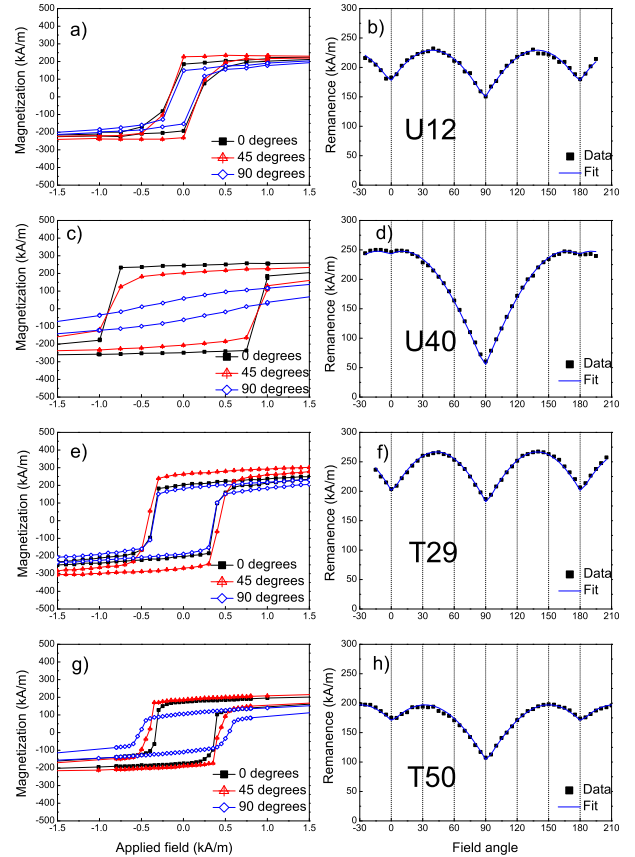}
\caption{[Color online] a) Magnetization loops with the field aligned with 3 high symmetry directions and b) remanence versus field angle dependence obtained from sample U12. The latter graph shows the data from the measurements (black dots) and the result from the fit procedure described in the text (blue line). c) and d) The corresponding graphs for sample U40. e) and f) Sample T29. g) and h) Sample T50. The field angle is defined with respect to the crystal structure, 0$^\circ$ (90$^\circ$) corresponds to the [001]$_\textrm{o}$ ([1$\overline{1}$0]$_\textrm{o}$) lattice direction.}
\label{lsat13hb86mag}
\end{figure}

The samples were characterized with vibrating sample magnetometer (VSM, Model 10 VSM by Microsense) measurements at room temperature. The in-plane angle of the applied field was varied to determine the magnetic anisotropy. For all field angles a full magnetization loop was measured and the remanent magnetization was obtained from the loop. Figure~\ref{lsat13hb86mag}a shows the magnetization loops of sample U12 with the field aligned with three high symmetry directions. In figure~\ref{lsat13hb86mag}b we plotted the dependence of the remanent magnetization on the in-plane field angle. The largest remanence was found for an applied field at approximately 40 degrees with respect to the [001]$_\textrm{o}$ in-plane lattice direction. A predominant biaxial behavior is observed with easy axes aligned with the [110]$_\textrm{pc}$ and symmetry related crystal directions. Next to this biaxial anisotropy a small uniaxial anisotropy is present as well which can be seen in the difference in remanent magnetization at 0 and 90 degrees and the shift of the easy axes to $\pm$40$^\circ$. This uniaxial contribution becomes more pronounced in the thicker film (sample U40) shown in figure \ref{lsat13hb86mag}c and \ref{lsat13hb86mag}d. The remanent magnetization of sample B at 0 degrees approaches the easy axes value and the remanent magnetization at 90 degrees is much smaller, only 20 percent of the easy axes value. Due to the combination of uniaxial and biaxial anisotropy the easy axes are shifted to $\pm$15$^\circ$. The hard axis magnetization loop does not show a switch of the magnetization, but an almost linear dependence of the magnetization on the applied field which is characteristic for a hard axis loop in a sample with a uniaxial anisotropy. 

A similar thickness dependence of the anisotropy was also found for the samples T29 and T50, which are twinned. The results are plotted in figure~\ref{lsat13hb86mag}e-\ref{lsat13hb86mag}h. A biaxial and a uniaxial contribution are present and the uniaxial contribution is more pronounced in the thicker film. Sample T50 shows easy axes which are shifted to $\pm$30$^\circ$ and the hard axis remanence is 50 percent of the easy axis value. Comparing samples U12 and U40 with samples T29 and T50 we find that the uniaxial contribution is more pronounced in the samples U12 and U40. 

\section{Discussion}

In order to quantify the biaxial and uniaxial contribution to the anisotropy we start with a general anisotropy energy equation which contains both a biaxial and a uniaxial contribution \cite{Chikazumi1964}.
\begin{equation}
E_a/V = k_\textrm{u} \sin^2(\phi-\phi_1) + \frac{k_1}{4} \sin^2(2(\phi-\phi_2)),
\label{energy}
\end{equation}
in which $E_a$ is the anisotropy energy, $V$ the volume of the sample, $k_\textrm{u}$ ($k_1$) the uniaxial (biaxial) anisotropy constant, $\phi$ the angle of the magnetization, $\phi_1$=0$^\circ$ ($\phi_2$=45$^\circ$) the angle of the easy axis of the uniaxial (biaxial) anisotropy. The easy axes are found by minimizing the energy with respect to $\phi$. This results in:
\begin{equation}
\left\{
\begin{array}{ll}
\cos(2\phi_{\textrm{easy}}) = k_\textrm{u}/k_1 & \textrm{for }k_\textrm{u}<k_1,\\
\phi_\textrm{easy}=0 & \textrm{for }k_\textrm{u}\geq k_1,\\
\end{array}\right.
\label{phieasy}
\end{equation}
from which the easy axes, $\phi_\textrm{easy}$, can be obtained. The measured remanence versus field angle dependencies are the projections of the magnetization, which at remanence is aligned with the easy axis, onto the measurement direction:
\begin{equation}
M_\textrm{rem} (\theta) = M_0 \cos(\theta-\phi_\textrm{easy}),
\label{rem}
\end{equation}
where $M_\textrm{rem}$ is the remanent magnetization, $\theta$ is the field angle, $M_0$ is the remanence in the easy direction and $\phi_\textrm{easy}$ is the closest easy axis. Equation \ref{rem} has been used to obtain the fits in figure \ref{lsat13hb86mag}. The measured data, and therefore the anisotropy, is well described by equation~\ref{energy}-\ref{rem}. This allows us to extract the ratio between the uniaxial and biaxial anisotropy energies $k_\textrm{u}/k_1$. The results are presented in table \ref{table}. For all samples the uniaxial anisotropy energy is found to be smaller than the biaxial anisotropy energy. Between the different samples the ratio $k_\textrm{u}/k_1$ changes by a factor of 10.

The anisotropy field $H_\textrm{an}$ can be obtained from the slope d$M$/d$H$ at $H$=0 in the hard axis magnetization loop of a material with uniaxial anisotropy by the relation
\begin{equation}
H_\textrm{an} = \frac{M_\textrm{sat}}{\textrm{d}M/\textrm{d}H}.
\label{ku1}
\end{equation}
Here, $M_\textrm{sat}$ is the saturation magnetization, $M$ is the magnetization and $H$ is the applied field. The anisotropy energy is given by
\begin{equation}
2 k_\textrm{u} = H_\textrm{an} \mu_0 M_\textrm{sat},
\label{ku2}
\end{equation}
in which $\mu_0$ is the permeability of free space. We obtained a value for $k_\textrm{u}$ of 540$\pm$5 J/m$^3$ for the almost uniaxial sample U40. This would imply a biaxial anisotropy constant of 600$\pm$10 J/m$^3$ which corresponds well with earlier obtained results \cite{Steenbeck1999, MathewsPHD}. Therefore one can assume that the $k_1$ value is the same for all films and independent of thickness. The uniaxial anisotropy constants are calculated from the $k_\textrm{u}/k_1$ ratios and presented in table~\ref{table} as well. The uniaxial anisotropy is very weak for these samples, only in the range 50-600 J/m$^3$.

Next we turn to the origin of both contributions to the anisotropy. The biaxial contribution with easy axes aligned with the $<$110$>$$_\textrm{pc}$ lattice directions corresponds well with earlier results of magnetic anisotropy of LSMO on STO (001)$_\textrm{c}$ substrates \cite{Steenbeck1999, Tsui2000}. This represents the intrinsic magnetocrystalline anisotropy of LSMO strained to in-plane cubic symmetry.  For the uniaxial contribution different explanations exist. It is well known that a weak uniaxial anisotropy in LSMO can be the result of stepedge induced anisotropy \cite{Wang2003,Mathews2005}. However, the thickness dependence of the uniaxial anisotropy is at odds with an interpretation in terms of stepedge induced anisotropy. The stepedge induced anisotropy compared to the biaxial anisotropy should scale with the ratio of the volume of the surface layers containing the stepedges (in practice the monolayers at the interfaces) to the volume of the film. Therefore the stepedge induced anisotropy should be most pronounced for the thinnest films as the miscut of the samples was comparable. Here the opposite is observed. Also the uniaxial easy axis was not aligned with the stepedge directions in most of the samples, sometimes even 90 degrees perpendicular to the stepedges. This rules out the contribution of stepedge induced anisotropy. 

The uniaxial easy axis is aligned with the [001]$_\textrm{o}$ lattice direction while the hard axis is aligned with the [1$\overline{1}$0]$_\textrm{o}$ lattice direction. This, together with the reduced uniaxial anisotropy for the twinned samples, relates the observed anisotropy to the orthorhombicity of the LSMO films. It is unclear what the origin of the magnetic anisotropy in the orthorhombic crystal structure is. In general magnetic anisotropy in manganites is explained by a global strain field which relates the easy axis to the maximum strain direction \cite{Tokuracmobookh4}. This cannot be applied for these samples, as the LSMO has equal strain in the [001]$_\textrm{o}$ and the [1$\overline{1}$0]$_\textrm{o}$ lattice directions due to the cubic symmetry of the substrate. Note that for the case of orthorhombic LSMO on NGO (110)$_\textrm{o}$ the uniaxial easy axis is aligned with the [1$\overline{1}$0]$_\textrm{o}$ lattice direction and is due to strain anisotropy \cite{Boschker2009}, in contrast to the observed anisotropy of LSMO on LSAT. 

The difference between the two lattice directions in the orthorhombic symmetry is due to the different oxygen octahedra rotations. We therefore suspect that the microscopic origin of the magnetocrystalline anisotropy in LSMO is effected by the oxygen octahedra rotations. A plausible scenario would be that the difference in octahedra rotations along the [1$\overline{1}$0]$_\textrm{o}$ and [001]$_\textrm{o}$ directions results in a different Mn-O-Mn bond angle ($<\textrm{Mn-O-Mn}>$)in these two directions. For manganites it is well known that the bandwidth, due to the double exchange mechanism, depends on the Mn-O-Mn bond angle according to \cite{Radaelli1997}
\begin{equation}
W=\frac{cos\omega}{d^{3.5}_{MnO}}
\end{equation}
in which $W$ is the bandwidth, $\omega=0.5(\pi-<\textrm{Mn-O-Mn}>$ is proportional to the amount of octahedra rotation and $d_{MnO}$ is the Mn-O distance. This suggests that for LSMO on LSAT the Mn-O-Mn bond angle is larger in the [001]$_\textrm{o}$ direction than in the [1$\overline{1}$0]$_\textrm{o}$ direction, resulting in an increase in the bandwidth and therefore the easy axis is aligned with this direction. This explanation is also consistent with the increase of $k_\textrm{u}/k_1$ with film thickness. Due to structural reconstructions at the interface and the surface \cite{Herger2008}, the oxygen octahedra rotations in the surface and interfacial regions deviate from those of the bulk of the film. The biaxial anisotropy is intrinsic to MnO$_6$ octahedra and not sensitive to the structural reconstructions. In a follow up work we will investigate the crystal structure, and especially the octahedra rotations, in more detail, in order to resolve the microscopic origin of the magnetocrystalline anisotropy in LSMO.

An alternative interpretation of the magnetic data is that the anisotropy is somehow related to the lattice modulations observed with the satellites in the XRD measurements. As the lattice modulations only occur in the orthorhombic [001]$_\textrm{o}$ direction, it is not possible to discriminate between anisotropy induced by the orthorhombic crystal structure and anisotropy induced by the lattice modulations in our experiment. Nevertheless, we expect that the lattice modulations result in microtwins which have reduced magnetic coupling at the microtwin boundaries. The shape anisotropy of each microtwin would then be aligned with the [1$\overline{1}$0]$_\textrm{o}$ direction, contrary to the observation of a magnetic [001]$_\textrm{o}$ easy axis. 

Although one would suspect that the uniaxial anisotropy would disappear for the twinned samples, this is not the case. We assume that this is due to the dominant presence of grains with one orientation.

\section{Conclusion}
LSMO films with an orthorhombic crystal structure can be grown coherently and untwinned on cubic LSAT substrates. The magnetic anisotropy of the films is described by a combination of biaxial anisotropy with easy axes along the $<$110$>$$_\textrm{pc}$ directions and a uniaxial anisotropy with easy axis along the [001]$_\textrm{o}$ direction. For thicker films the uniaxial anisotropy becomes more pronounced. The uniaxial part of the anisotropy is induced by the orthorhombic symmetry of the LSMO. We expect these findings to lead to a better understanding of the microscopic origin of the magnetocrystalline anisotropy in LSMO. 

This research was financially supported by the Dutch Science Foundation, by NanoNed, a nanotechnology program of the Dutch Ministry of Economic Affairs and by the NanOxide program of the European Science Foundation.

\bibliographystyle{apsrev}

\begin{thebibliography}{31}
\expandafter\ifx\csname natexlab\endcsname\relax\def\natexlab#1{#1}\fi
\expandafter\ifx\csname bibnamefont\endcsname\relax
  \def\bibnamefont#1{#1}\fi
\expandafter\ifx\csname bibfnamefont\endcsname\relax
  \def\bibfnamefont#1{#1}\fi
\expandafter\ifx\csname citenamefont\endcsname\relax
  \def\citenamefont#1{#1}\fi
\expandafter\ifx\csname url\endcsname\relax
  \def\url#1{\texttt{#1}}\fi
\expandafter\ifx\csname urlprefix\endcsname\relax\def\urlprefix{URL }\fi
\providecommand{\bibinfo}[2]{#2}
\providecommand{\eprint}[2][]{\url{#2}}

\bibitem[{\citenamefont{Vonhelmolt et~al.}(1993)\citenamefont{Vonhelmolt,
  Wecker, Holzapfel, Schultz, and Samwer}}]{Helmolt1993}
\bibinfo{author}{\bibfnamefont{R.}~\bibnamefont{Vonhelmolt}},
  \bibinfo{author}{\bibfnamefont{J.}~\bibnamefont{Wecker}},
  \bibinfo{author}{\bibfnamefont{B.}~\bibnamefont{Holzapfel}},
  \bibinfo{author}{\bibfnamefont{L.}~\bibnamefont{Schultz}}, \bibnamefont{and}
  \bibinfo{author}{\bibfnamefont{K.}~\bibnamefont{Samwer}},
  \bibinfo{journal}{PHYSICAL REVIEW LETTERS} \textbf{\bibinfo{volume}{71}},
  \bibinfo{pages}{2331} (\bibinfo{year}{1993}), ISSN \bibinfo{issn}{0031-9007}.

\bibitem[{\citenamefont{Jin et~al.}(1994)\citenamefont{Jin, Tiefel, Mccormack,
  Fastnacht, Ramesh, and Chen}}]{Jin1994}
\bibinfo{author}{\bibfnamefont{S.}~\bibnamefont{Jin}},
  \bibinfo{author}{\bibfnamefont{T.}~\bibnamefont{Tiefel}},
  \bibinfo{author}{\bibfnamefont{M.}~\bibnamefont{Mccormack}},
  \bibinfo{author}{\bibfnamefont{R.}~\bibnamefont{Fastnacht}},
  \bibinfo{author}{\bibfnamefont{R.}~\bibnamefont{Ramesh}}, \bibnamefont{and}
  \bibinfo{author}{\bibfnamefont{L.}~\bibnamefont{Chen}},
  \bibinfo{journal}{SCIENCE} \textbf{\bibinfo{volume}{264}},
  \bibinfo{pages}{413} (\bibinfo{year}{1994}), ISSN \bibinfo{issn}{0036-8075}.

\bibitem[{\citenamefont{Moritomo et~al.}(1995)\citenamefont{Moritomo, Asamitsu,
  and Tokura}}]{Moritomo1995}
\bibinfo{author}{\bibfnamefont{Y.}~\bibnamefont{Moritomo}},
  \bibinfo{author}{\bibfnamefont{A.}~\bibnamefont{Asamitsu}}, \bibnamefont{and}
  \bibinfo{author}{\bibfnamefont{Y.}~\bibnamefont{Tokura}},
  \bibinfo{journal}{PHYSICAL REVIEW B} \textbf{\bibinfo{volume}{51}},
  \bibinfo{pages}{16491} (\bibinfo{year}{1995}), ISSN
  \bibinfo{issn}{0163-1829}.

\bibitem[{\citenamefont{Hwang et~al.}(1995{\natexlab{a}})\citenamefont{Hwang,
  Palstra, Cheong, and Batlogg}}]{Hwang1995prb}
\bibinfo{author}{\bibfnamefont{H.}~\bibnamefont{Hwang}},
  \bibinfo{author}{\bibfnamefont{T.}~\bibnamefont{Palstra}},
  \bibinfo{author}{\bibfnamefont{S.}~\bibnamefont{Cheong}}, \bibnamefont{and}
  \bibinfo{author}{\bibfnamefont{B.}~\bibnamefont{Batlogg}},
  \bibinfo{journal}{PHYSICAL REVIEW B} \textbf{\bibinfo{volume}{52}},
  \bibinfo{pages}{15046} (\bibinfo{year}{1995}{\natexlab{a}}), ISSN
  \bibinfo{issn}{0163-1829}.

\bibitem[{\citenamefont{Park et~al.}(1998)\citenamefont{Park, Vescovo, Kim,
  Kwon, Ramesh, and Venkatesan}}]{Park1998}
\bibinfo{author}{\bibfnamefont{J.}~\bibnamefont{Park}},
  \bibinfo{author}{\bibfnamefont{E.}~\bibnamefont{Vescovo}},
  \bibinfo{author}{\bibfnamefont{H.}~\bibnamefont{Kim}},
  \bibinfo{author}{\bibfnamefont{C.}~\bibnamefont{Kwon}},
  \bibinfo{author}{\bibfnamefont{R.}~\bibnamefont{Ramesh}}, \bibnamefont{and}
  \bibinfo{author}{\bibfnamefont{T.}~\bibnamefont{Venkatesan}},
  \bibinfo{journal}{NATURE} \textbf{\bibinfo{volume}{392}},
  \bibinfo{pages}{794} (\bibinfo{year}{1998}), ISSN \bibinfo{issn}{0028-0836}.

\bibitem[{\citenamefont{Bowen et~al.}(2003)\citenamefont{Bowen, Bibes,
  Barthelemy, Contour, Anane, Lemaitre, and Fert}}]{Bowen2003}
\bibinfo{author}{\bibfnamefont{M.}~\bibnamefont{Bowen}},
  \bibinfo{author}{\bibfnamefont{M.}~\bibnamefont{Bibes}},
  \bibinfo{author}{\bibfnamefont{A.}~\bibnamefont{Barthelemy}},
  \bibinfo{author}{\bibfnamefont{J.}~\bibnamefont{Contour}},
  \bibinfo{author}{\bibfnamefont{A.}~\bibnamefont{Anane}},
  \bibinfo{author}{\bibfnamefont{Y.}~\bibnamefont{Lemaitre}}, \bibnamefont{and}
  \bibinfo{author}{\bibfnamefont{A.}~\bibnamefont{Fert}},
  \bibinfo{journal}{APPLIED PHYSICS LETTERS} \textbf{\bibinfo{volume}{82}},
  \bibinfo{pages}{233} (\bibinfo{year}{2003}), ISSN \bibinfo{issn}{0003-6951}.

\bibitem[{\citenamefont{Kwon et~al.}(1997)\citenamefont{Kwon, Robson, Kim, Gu,
  Lofland, Bhagat, Trajanovic, Rajeswari, Venkatesan, Kratz et~al.}}]{Kwon1997}
\bibinfo{author}{\bibfnamefont{C.}~\bibnamefont{Kwon}},
  \bibinfo{author}{\bibfnamefont{M.}~\bibnamefont{Robson}},
  \bibinfo{author}{\bibfnamefont{K.}~\bibnamefont{Kim}},
  \bibinfo{author}{\bibfnamefont{J.}~\bibnamefont{Gu}},
  \bibinfo{author}{\bibfnamefont{S.}~\bibnamefont{Lofland}},
  \bibinfo{author}{\bibfnamefont{S.}~\bibnamefont{Bhagat}},
  \bibinfo{author}{\bibfnamefont{Z.}~\bibnamefont{Trajanovic}},
  \bibinfo{author}{\bibfnamefont{M.}~\bibnamefont{Rajeswari}},
  \bibinfo{author}{\bibfnamefont{T.}~\bibnamefont{Venkatesan}},
  \bibinfo{author}{\bibfnamefont{A.}~\bibnamefont{Kratz}},
  \bibnamefont{et~al.}, \bibinfo{journal}{JOURNAL OF MAGNETISM AND MAGNETIC
  MATERIALS} \textbf{\bibinfo{volume}{172}}, \bibinfo{pages}{229}
  (\bibinfo{year}{1997}), ISSN \bibinfo{issn}{0304-8853}.

\bibitem[{\citenamefont{Suzuki et~al.}(1998)\citenamefont{Suzuki, Hwang,
  Cheong, Siegrist, van Dover, Asamitsu, and Tokura}}]{Suzuki1998}
\bibinfo{author}{\bibfnamefont{Y.}~\bibnamefont{Suzuki}},
  \bibinfo{author}{\bibfnamefont{H.}~\bibnamefont{Hwang}},
  \bibinfo{author}{\bibfnamefont{S.}~\bibnamefont{Cheong}},
  \bibinfo{author}{\bibfnamefont{T.}~\bibnamefont{Siegrist}},
  \bibinfo{author}{\bibfnamefont{R.}~\bibnamefont{van Dover}},
  \bibinfo{author}{\bibfnamefont{A.}~\bibnamefont{Asamitsu}}, \bibnamefont{and}
  \bibinfo{author}{\bibfnamefont{Y.}~\bibnamefont{Tokura}},
  \bibinfo{journal}{JOURNAL OF APPLIED PHYSICS} \textbf{\bibinfo{volume}{83}},
  \bibinfo{pages}{7064} (\bibinfo{year}{1998}), ISSN \bibinfo{issn}{0021-8979}.

\bibitem[{\citenamefont{Steenbeck and Hiergeist}(1999)}]{Steenbeck1999}
\bibinfo{author}{\bibfnamefont{K.}~\bibnamefont{Steenbeck}} \bibnamefont{and}
  \bibinfo{author}{\bibfnamefont{R.}~\bibnamefont{Hiergeist}},
  \bibinfo{journal}{APPLIED PHYSICS LETTERS} \textbf{\bibinfo{volume}{75}},
  \bibinfo{pages}{1778} (\bibinfo{year}{1999}), ISSN \bibinfo{issn}{0003-6951}.

\bibitem[{\citenamefont{Tsui et~al.}(2000)\citenamefont{Tsui, Smoak, Nath, and
  Eom}}]{Tsui2000}
\bibinfo{author}{\bibfnamefont{F.}~\bibnamefont{Tsui}},
  \bibinfo{author}{\bibfnamefont{M.}~\bibnamefont{Smoak}},
  \bibinfo{author}{\bibfnamefont{T.}~\bibnamefont{Nath}}, \bibnamefont{and}
  \bibinfo{author}{\bibfnamefont{C.}~\bibnamefont{Eom}},
  \bibinfo{journal}{APPLIED PHYSICS LETTERS} \textbf{\bibinfo{volume}{76}},
  \bibinfo{pages}{2421} (\bibinfo{year}{2000}), ISSN \bibinfo{issn}{0003-6951}.

\bibitem[{\citenamefont{Desfeux et~al.}(2001)\citenamefont{Desfeux, Bailleul,
  Da~Costa, Prellier, and Haghiri-Gosnet}}]{Desfeux2001}
\bibinfo{author}{\bibfnamefont{R.}~\bibnamefont{Desfeux}},
  \bibinfo{author}{\bibfnamefont{S.}~\bibnamefont{Bailleul}},
  \bibinfo{author}{\bibfnamefont{A.}~\bibnamefont{Da~Costa}},
  \bibinfo{author}{\bibfnamefont{W.}~\bibnamefont{Prellier}}, \bibnamefont{and}
  \bibinfo{author}{\bibfnamefont{A.}~\bibnamefont{Haghiri-Gosnet}},
  \bibinfo{journal}{APPLIED PHYSICS LETTERS} \textbf{\bibinfo{volume}{78}},
  \bibinfo{pages}{3681} (\bibinfo{year}{2001}), ISSN \bibinfo{issn}{0003-6951}.

\bibitem[{\citenamefont{Dho et~al.}(2003)\citenamefont{Dho, Kim, Hwang, Kim,
  and Hur}}]{Dho2003}
\bibinfo{author}{\bibfnamefont{J.}~\bibnamefont{Dho}},
  \bibinfo{author}{\bibfnamefont{Y.}~\bibnamefont{Kim}},
  \bibinfo{author}{\bibfnamefont{Y.}~\bibnamefont{Hwang}},
  \bibinfo{author}{\bibfnamefont{J.}~\bibnamefont{Kim}}, \bibnamefont{and}
  \bibinfo{author}{\bibfnamefont{N.}~\bibnamefont{Hur}},
  \bibinfo{journal}{APPLIED PHYSICS LETTERS} \textbf{\bibinfo{volume}{82}},
  \bibinfo{pages}{1434} (\bibinfo{year}{2003}), ISSN \bibinfo{issn}{0003-6951}.

\bibitem[{\citenamefont{Mathews et~al.}(2005)\citenamefont{Mathews, Postma,
  Lodder, Jansen, Rijnders, and Blank}}]{Mathews2005}
\bibinfo{author}{\bibfnamefont{M.}~\bibnamefont{Mathews}},
  \bibinfo{author}{\bibfnamefont{F.}~\bibnamefont{Postma}},
  \bibinfo{author}{\bibfnamefont{J.}~\bibnamefont{Lodder}},
  \bibinfo{author}{\bibfnamefont{R.}~\bibnamefont{Jansen}},
  \bibinfo{author}{\bibfnamefont{G.}~\bibnamefont{Rijnders}}, \bibnamefont{and}
  \bibinfo{author}{\bibfnamefont{D.}~\bibnamefont{Blank}},
  \bibinfo{journal}{APPLIED PHYSICS LETTERS} \textbf{\bibinfo{volume}{87}}
  (\bibinfo{year}{2005}), ISSN \bibinfo{issn}{0003-6951}.

\bibitem[{\citenamefont{Boschker et~al.}(2009)\citenamefont{Boschker, Mathews,
  Houwman, Nishikawa, Vailionis, Koster, Rijnders, and Blank}}]{Boschker2009}
\bibinfo{author}{\bibfnamefont{H.}~\bibnamefont{Boschker}},
  \bibinfo{author}{\bibfnamefont{M.}~\bibnamefont{Mathews}},
  \bibinfo{author}{\bibfnamefont{E.~P.} \bibnamefont{Houwman}},
  \bibinfo{author}{\bibfnamefont{H.}~\bibnamefont{Nishikawa}},
  \bibinfo{author}{\bibfnamefont{A.}~\bibnamefont{Vailionis}},
  \bibinfo{author}{\bibfnamefont{G.}~\bibnamefont{Koster}},
  \bibinfo{author}{\bibfnamefont{G.}~\bibnamefont{Rijnders}}, \bibnamefont{and}
  \bibinfo{author}{\bibfnamefont{D.~H.~A.} \bibnamefont{Blank}},
  \bibinfo{journal}{PHYSICAL REVIEW B} \textbf{\bibinfo{volume}{79}}
  (\bibinfo{year}{2009}), ISSN \bibinfo{issn}{1098-0121}.

\bibitem[{\citenamefont{Mathews et~al.}(2010)\citenamefont{Mathews, Houwman,
  Boschker, Rijnders, and Blank}}]{Mathews2010}
\bibinfo{author}{\bibfnamefont{M.}~\bibnamefont{Mathews}},
  \bibinfo{author}{\bibfnamefont{E.~P.} \bibnamefont{Houwman}},
  \bibinfo{author}{\bibfnamefont{H.}~\bibnamefont{Boschker}},
  \bibinfo{author}{\bibfnamefont{G.}~\bibnamefont{Rijnders}}, \bibnamefont{and}
  \bibinfo{author}{\bibfnamefont{D.~H.~A.} \bibnamefont{Blank}},
  \bibinfo{journal}{JOURNAL OF APPLIED PHYSICS} \textbf{\bibinfo{volume}{107}}
  (\bibinfo{year}{2010}).

\bibitem[{\citenamefont{Wang et~al.}(2003)\citenamefont{Wang, Cristiani, and
  Habermeier}}]{Wang2003}
\bibinfo{author}{\bibfnamefont{Z.}~\bibnamefont{Wang}},
  \bibinfo{author}{\bibfnamefont{G.}~\bibnamefont{Cristiani}},
  \bibnamefont{and}
  \bibinfo{author}{\bibfnamefont{H.}~\bibnamefont{Habermeier}},
  \bibinfo{journal}{APPLIED PHYSICS LETTERS} \textbf{\bibinfo{volume}{82}},
  \bibinfo{pages}{3731} (\bibinfo{year}{2003}), ISSN \bibinfo{issn}{0003-6951}.

\bibitem[{\citenamefont{Vailionis et~al.}(2008)\citenamefont{Vailionis,
  Siemons, and Koster}}]{Vailionis2008}
\bibinfo{author}{\bibfnamefont{A.}~\bibnamefont{Vailionis}},
  \bibinfo{author}{\bibfnamefont{W.}~\bibnamefont{Siemons}}, \bibnamefont{and}
  \bibinfo{author}{\bibfnamefont{G.}~\bibnamefont{Koster}},
  \bibinfo{journal}{APPLIED PHYSICS LETTERS} \textbf{\bibinfo{volume}{93}}
  (\bibinfo{year}{2008}), ISSN \bibinfo{issn}{0003-6951}.

\bibitem[{\citenamefont{Vailionis et~al.}(2009)\citenamefont{Vailionis,
  Boschker, Houwman, Koster, Rijnders, and Blank}}]{Vailionis2009}
\bibinfo{author}{\bibfnamefont{A.}~\bibnamefont{Vailionis}},
  \bibinfo{author}{\bibfnamefont{H.}~\bibnamefont{Boschker}},
  \bibinfo{author}{\bibfnamefont{E.}~\bibnamefont{Houwman}},
  \bibinfo{author}{\bibfnamefont{G.}~\bibnamefont{Koster}},
  \bibinfo{author}{\bibfnamefont{G.}~\bibnamefont{Rijnders}}, \bibnamefont{and}
  \bibinfo{author}{\bibfnamefont{D.~H.~A.} \bibnamefont{Blank}},
  \bibinfo{journal}{APPLIED PHYSICS LETTERS} \textbf{\bibinfo{volume}{95}}
  (\bibinfo{year}{2009}), ISSN \bibinfo{issn}{0003-6951}.

\bibitem[{\citenamefont{Hwang et~al.}(1995{\natexlab{b}})\citenamefont{Hwang,
  Cheong, Radaelli, Marezio, and Batlogg}}]{Hwang1995}
\bibinfo{author}{\bibfnamefont{H.}~\bibnamefont{Hwang}},
  \bibinfo{author}{\bibfnamefont{S.}~\bibnamefont{Cheong}},
  \bibinfo{author}{\bibfnamefont{P.}~\bibnamefont{Radaelli}},
  \bibinfo{author}{\bibfnamefont{M.}~\bibnamefont{Marezio}}, \bibnamefont{and}
  \bibinfo{author}{\bibfnamefont{B.}~\bibnamefont{Batlogg}},
  \bibinfo{journal}{PHYSICAL REVIEW LETTERS} \textbf{\bibinfo{volume}{75}},
  \bibinfo{pages}{914} (\bibinfo{year}{1995}{\natexlab{b}}), ISSN
  \bibinfo{issn}{0031-9007}.

\bibitem[{\citenamefont{Radaelli et~al.}(1997)\citenamefont{Radaelli, Iannone,
  Marezio, Hwang, Cheong, Jorgensen, and Argyriou}}]{Radaelli1997}
\bibinfo{author}{\bibfnamefont{P.}~\bibnamefont{Radaelli}},
  \bibinfo{author}{\bibfnamefont{G.}~\bibnamefont{Iannone}},
  \bibinfo{author}{\bibfnamefont{M.}~\bibnamefont{Marezio}},
  \bibinfo{author}{\bibfnamefont{H.}~\bibnamefont{Hwang}},
  \bibinfo{author}{\bibfnamefont{S.}~\bibnamefont{Cheong}},
  \bibinfo{author}{\bibfnamefont{J.}~\bibnamefont{Jorgensen}},
  \bibnamefont{and} \bibinfo{author}{\bibfnamefont{D.}~\bibnamefont{Argyriou}},
  \bibinfo{journal}{PHYSICAL REVIEW B} \textbf{\bibinfo{volume}{56}},
  \bibinfo{pages}{8265} (\bibinfo{year}{1997}), ISSN \bibinfo{issn}{0163-1829}.

\bibitem[{\citenamefont{Mathews}(2007)}]{MathewsPHD}
\bibinfo{author}{\bibfnamefont{M.}~\bibnamefont{Mathews}}, Ph.D. thesis,
  \bibinfo{school}{University of Twente} (\bibinfo{year}{2007}).

\bibitem[{\citenamefont{Vigliante et~al.}(2001)\citenamefont{Vigliante,
  Gebhardt, Ruhm, Wochner, Razavi, and Habermeier}}]{Vigliante2001}
\bibinfo{author}{\bibfnamefont{A.}~\bibnamefont{Vigliante}},
  \bibinfo{author}{\bibfnamefont{U.}~\bibnamefont{Gebhardt}},
  \bibinfo{author}{\bibfnamefont{A.}~\bibnamefont{Ruhm}},
  \bibinfo{author}{\bibfnamefont{P.}~\bibnamefont{Wochner}},
  \bibinfo{author}{\bibfnamefont{F.}~\bibnamefont{Razavi}}, \bibnamefont{and}
  \bibinfo{author}{\bibfnamefont{H.}~\bibnamefont{Habermeier}},
  \bibinfo{journal}{EUROPHYSICS LETTERS} \textbf{\bibinfo{volume}{54}},
  \bibinfo{pages}{619} (\bibinfo{year}{2001}), ISSN \bibinfo{issn}{0295-5075}.

\bibitem[{\citenamefont{Farag et~al.}(2005)\citenamefont{Farag, Bobeth, Pompe,
  Romanov, and Speck}}]{Farag2005}
\bibinfo{author}{\bibfnamefont{N.}~\bibnamefont{Farag}},
  \bibinfo{author}{\bibfnamefont{M.}~\bibnamefont{Bobeth}},
  \bibinfo{author}{\bibfnamefont{W.}~\bibnamefont{Pompe}},
  \bibinfo{author}{\bibfnamefont{A.}~\bibnamefont{Romanov}}, \bibnamefont{and}
  \bibinfo{author}{\bibfnamefont{J.}~\bibnamefont{Speck}},
  \bibinfo{journal}{JOURNAL OF APPLIED PHYSICS} \textbf{\bibinfo{volume}{97}}
  (\bibinfo{year}{2005}), ISSN \bibinfo{issn}{0021-8979}.

\bibitem[{\citenamefont{Gebhardt et~al.}(2007)\citenamefont{Gebhardt, Kasper,
  Vigliante, Wochner, Dosch, Razavi, and Habermeier}}]{Gebhardt2007}
\bibinfo{author}{\bibfnamefont{U.}~\bibnamefont{Gebhardt}},
  \bibinfo{author}{\bibfnamefont{N.~V.} \bibnamefont{Kasper}},
  \bibinfo{author}{\bibfnamefont{A.}~\bibnamefont{Vigliante}},
  \bibinfo{author}{\bibfnamefont{P.}~\bibnamefont{Wochner}},
  \bibinfo{author}{\bibfnamefont{H.}~\bibnamefont{Dosch}},
  \bibinfo{author}{\bibfnamefont{F.~S.} \bibnamefont{Razavi}},
  \bibnamefont{and} \bibinfo{author}{\bibfnamefont{H.~U.}
  \bibnamefont{Habermeier}}, \bibinfo{journal}{PHYSICAL REVIEW LETTERS}
  \textbf{\bibinfo{volume}{98}} (\bibinfo{year}{2007}), ISSN
  \bibinfo{issn}{0031-9007}.

\bibitem[{\citenamefont{Zhou et~al.}(2007)\citenamefont{Zhou, Li, Li, Jin, and
  Wu}}]{Zhou2007}
\bibinfo{author}{\bibfnamefont{T.~F.} \bibnamefont{Zhou}},
  \bibinfo{author}{\bibfnamefont{G.}~\bibnamefont{Li}},
  \bibinfo{author}{\bibfnamefont{X.~G.} \bibnamefont{Li}},
  \bibinfo{author}{\bibfnamefont{S.~W.} \bibnamefont{Jin}}, \bibnamefont{and}
  \bibinfo{author}{\bibfnamefont{W.~B.} \bibnamefont{Wu}},
  \bibinfo{journal}{APPLIED PHYSICS LETTERS} \textbf{\bibinfo{volume}{90}}
  (\bibinfo{year}{2007}), ISSN \bibinfo{issn}{0003-6951}.

\bibitem[{\citenamefont{Gan et~al.}(1997)\citenamefont{Gan, Rao, and
  Eom}}]{Gan1997}
\bibinfo{author}{\bibfnamefont{Q.}~\bibnamefont{Gan}},
  \bibinfo{author}{\bibfnamefont{R.}~\bibnamefont{Rao}}, \bibnamefont{and}
  \bibinfo{author}{\bibfnamefont{C.}~\bibnamefont{Eom}},
  \bibinfo{journal}{APPLIED PHYSICS LETTERS} \textbf{\bibinfo{volume}{70}},
  \bibinfo{pages}{1962} (\bibinfo{year}{1997}), ISSN \bibinfo{issn}{0003-6951}.

\bibitem[{\citenamefont{Maria et~al.}(2000)\citenamefont{Maria, McKinstry, and
  Trolier-McKinstry}}]{Maria2000}
\bibinfo{author}{\bibfnamefont{J.}~\bibnamefont{Maria}},
  \bibinfo{author}{\bibfnamefont{H.}~\bibnamefont{McKinstry}},
  \bibnamefont{and}
  \bibinfo{author}{\bibfnamefont{S.}~\bibnamefont{Trolier-McKinstry}},
  \bibinfo{journal}{APPLIED PHYSICS LETTERS} \textbf{\bibinfo{volume}{76}},
  \bibinfo{pages}{3382} (\bibinfo{year}{2000}), ISSN \bibinfo{issn}{0003-6951}.

\bibitem[{\citenamefont{Vailionis et~al.}(2007)\citenamefont{Vailionis,
  Siemons, and Koster}}]{Vailionis2007}
\bibinfo{author}{\bibfnamefont{A.}~\bibnamefont{Vailionis}},
  \bibinfo{author}{\bibfnamefont{W.}~\bibnamefont{Siemons}}, \bibnamefont{and}
  \bibinfo{author}{\bibfnamefont{G.}~\bibnamefont{Koster}},
  \bibinfo{journal}{APPLIED PHYSICS LETTERS} \textbf{\bibinfo{volume}{91}}
  (\bibinfo{year}{2007}), ISSN \bibinfo{issn}{0003-6951}.

\bibitem[{\citenamefont{Chikazumi}(1964)}]{Chikazumi1964}
\bibinfo{author}{\bibfnamefont{S.}~\bibnamefont{Chikazumi}},
  \emph{\bibinfo{title}{Physics of ferromagnetism}} (\bibinfo{publisher}{John
  Wiley and Sons, Inc, New York}, \bibinfo{year}{1964}).

\bibitem[{\citenamefont{Terakura et~al.}(2000)\citenamefont{Terakura, Solovyev,
  and Sawada}}]{Tokuracmobookh4}
\bibinfo{author}{\bibfnamefont{K.}~\bibnamefont{Terakura}},
  \bibinfo{author}{\bibfnamefont{I.}~\bibnamefont{Solovyev}}, \bibnamefont{and}
  \bibinfo{author}{\bibfnamefont{H.}~\bibnamefont{Sawada}}, in
  \emph{\bibinfo{booktitle}{Colossal magnetoresistive oxides}}, edited by
  \bibinfo{editor}{\bibfnamefont{Y.}~\bibnamefont{Tokura}}
  (\bibinfo{publisher}{Gordon and Breach Science Publishers, the Netherlands},
  \bibinfo{year}{2000}), pp. \bibinfo{pages}{119--148}.

\bibitem[{\citenamefont{Herger et~al.}(2008)\citenamefont{Herger, Willmott,
  Schlepuetz, Bjoerck, Pauli, Martoccia, Patterson, Kumah, Clarke, Yacoby
  et~al.}}]{Herger2008}
\bibinfo{author}{\bibfnamefont{R.}~\bibnamefont{Herger}},
  \bibinfo{author}{\bibfnamefont{P.~R.} \bibnamefont{Willmott}},
  \bibinfo{author}{\bibfnamefont{C.~M.} \bibnamefont{Schlepuetz}},
  \bibinfo{author}{\bibfnamefont{M.}~\bibnamefont{Bjoerck}},
  \bibinfo{author}{\bibfnamefont{S.~A.} \bibnamefont{Pauli}},
  \bibinfo{author}{\bibfnamefont{D.}~\bibnamefont{Martoccia}},
  \bibinfo{author}{\bibfnamefont{B.~D.} \bibnamefont{Patterson}},
  \bibinfo{author}{\bibfnamefont{D.}~\bibnamefont{Kumah}},
  \bibinfo{author}{\bibfnamefont{R.}~\bibnamefont{Clarke}},
  \bibinfo{author}{\bibfnamefont{Y.}~\bibnamefont{Yacoby}},
  \bibnamefont{et~al.}, \bibinfo{journal}{PHYSICAL REVIEW B}
  \textbf{\bibinfo{volume}{77}} (\bibinfo{year}{2008}), ISSN
  \bibinfo{issn}{1098-0121}.

\end{thebibliography}

\end{document}